\documentclass[preprint,12pt,aps,prl]{revtex4-1}

\usepackage{graphicx}
\usepackage{amsmath,amssymb,latexsym}

\usepackage{times}
\usepackage{mathptmx}

\newcommand{\ibid}{\textit{ibid.}}

\newcommand{\wt}[1]{{\widetilde#1}}

\newcommand{\nG}{\ensuremath{n_{\text{G}}}}

\newcommand{\nebar}{\ensuremath{\overline{n}_\text{e}}}
\newcommand{\Ip}{\ensuremath{I_\text{p}}}

\newcommand{\taud}{\ensuremath{\tau_\text{d}}}
\newcommand{\tauw}{\ensuremath{\tau_\text{w}}}
\newcommand{\Phirms}{\ensuremath{\Phi_\text{rms}}}
\newcommand{\Phiave}{\ensuremath{\langle{\Phi}\rangle}}

\newcommand{\Aave}{\ensuremath{\langle{A}\rangle}}

\newcommand{\Ref}[1]{Ref.~\onlinecite{#1}}
\newcommand{\Refs}[1]{Refs.~\onlinecite{#1}}
\newcommand{\Eqref}[1]{Eq.~\eqref{#1}}

\newcommand{\Figref}[1]{Fig.~\ref{#1}}
\newcommand{\Figsref}[1]{Figs.~\ref{#1}}

\newcommand{\JNM}{\textit{J.~Nuclear Mater.}}

\newcommand{\NF}{\textit{Nucl.\ Fusion}}
\newcommand{\NME}{\textit{Nucl.\ Mater.\ Energy}}

\newcommand{\PPCF}{\textit{Plasma Phys.\ Contr.\ Fusion}}

\newcommand{\PP}{\textit{Phys.\ Plasmas}}

\newcommand{\PS}{\textit{Phys.\ Scripta}}
\newcommand{\RSI}{\textit{Rev.\ Sci.\ Instr.}}

\newcommand{\PRL}{\textit{Phys.~Rev.\ Lett.}}

\begin{document}

\title{Relationship between frequency power spectra and intermittent, large-amplitude bursts in the Alcator C-Mod scrape-off layer}
\date{\today}
\author{A.~Theodorsen}
\author{O.~E.~Garcia}
\email{odd.erik.garcia@uit.no}
\author{R.~Kube}
\affiliation{Department of Physics and Technology, UiT The Arctic University of Norway, N-9037 Troms{\o}, Norway}
\author{B.~La{B}ombard}
\author{J.~L.~Terry}
\affiliation{MIT Plasma Science and Fusion Center, Cambridge 02139, MA, USA}

\begin{abstract}
Fluctuations in the boundary region of the Alcator C-Mod tokamak have been analyzed using gas puff imaging data from a set of Ohmically heated plasma density scan experiments. It is found that the relative fluctuation amplitudes are modest and close to normally distributed at the separatrix but become increasingly larger and intermittent towards the main chamber wall. The frequency power spectra are nevertheless similar for all radial positions and line-averaged densities. Predictions of a stochastic model, describing the plasma fluctuations as a super-position of uncorrelated pulses, are shown to be in excellent agreement with the measurements. This implies that the pulse duration is the same, while the degree of pulse overlap decreases radially outwards in the scrape-off layer. The model also describes the rate of threshold level crossings, which provide novel predictions of plasma--wall interactions due to transient transport events.
\end{abstract}

\maketitle

Turbulent motions and fluctuation-induced transport of particles and heat in the boundary region of magnetically confined plasmas pose numerous problems for the successful development of a thermonuclear fusion power reactor \cite{dmz,lipschultz,labombard-2005}. This includes challenges for auxiliary heating and current drive by the use of radio frequency waves \cite{terry-lhrf,myra,wallace} and enhanced material erosion of the main chamber walls \cite{marandet,birkenmeier,dm-pwi}, and is likely related to the empirical discharge density limit \cite{greenwald,dm,guzdar,xu,labombard-2001,garcia-tcv}. At the outboard mid-plane scrape-off layer the radial transport is predominantly due to radially outwards motion of blob-like plasma filaments, which leads to order unity relative fluctuation levels, broad plasma profiles and enhanced plasma--wall interactions \cite{labombard-2001,garcia-tcv,labombard-2011,cziegler,garcia-aps,theodorsen-ppcf,kube-ppcf,garcia-nme}.

The statistical properties of plasma fluctuations in the scrape-off layer have recently been elucidated by means of exceptionally long data time series under stationary plasma conditions \cite{garcia-aps,theodorsen-ppcf,kube-ppcf,garcia-nme}. Based on this, a stochastic model of the plasma fluctuations has been developed \cite{garcia-php,theodorsen-php,garcia-psd}
and its underlying assumptions and predictions are found to compare favorably with experimental measurements \cite{garcia-aps,theodorsen-ppcf,kube-ppcf,garcia-nme}. Recently, analytical expressions for the rate of level crossings were obtained from this model, that is, how frequently a realization of the process on average crosses a given threshold level. This gives novel predictions of threshold phenomena underlying plasma--wall interactions due to transient transport events \cite{garcia-php,theodorsen-php}.

In this contribution, fluctuations in the boundary region of the Alcator C-Mod tokamak are investigated by analysis of gas puff imaging measurements. The present work may be considered as a continuation of the first studies of these measurement data published in \Ref{garcia-aps}. A major new finding is that the frequency power spectral density is similar for all radial positions and line-averaged densities. The spectrum agrees with the model prediction, implying that the temporal scale of the fluctuations is the same, while the degree of pulse overlap decreases radially outwards. Predictions of level crossing rates are also found to be in excellent agreement with the experimental measurements. The stochastic model thus describes the plasma fluctuations across the entire outboard mid-plane scrape-off layer in these Alcator C-Mod plasmas.

Gas puff imaging data from the boundary region of the Alcator C-Mod tokamak in Ohmically heated, lower single null discharges have been analyzed. This comprises a four-point scan in the line-averaged density $\nebar$, with the Greenwald fraction $\nebar/\nG$ ranging from $0.15$ to $0.30$. Here the Greenwald density is given by $\nG=(\Ip/\pi a^2)\,10^{20}\,\text{m}^{-3}$, where $\Ip$ is the plasma current in units of MA and $a$ is the plasma minor radius in units of meters. For this density scan $\Ip=0.8\,\text{MA}$ and $a=0.21\,\text{m}$, giving $\nG=5.26\times10^{20}\,\text{m}^{-3}$. An on-axis toroidal magnetic field of $4.0\,\text{T}$ was used. The condition at the outer divertor target goes from sheath limited at the lowest density to high recycling at the highest density in this density scan, and the particle density profile at the outboard mid-plane develops the familiar two-layer structure and significant broadening with increasing line-averaged density. These experiments were part of heat flux footprint studies on Alcator C-Mod, and have been extensively diagnosed and documented \cite{labombard-2011,garcia-aps}.

The gas puff imaging diagnostic consists of a $9\times10$ (radial$\,\times\,$vertical) toroidal views, coupled to an in-vessel optical fiber array. The plasma emission collected in the views is filtered for He I ($587\,$nm) line emission that is due almost entirely to plasma excitation near the object plane of an extended He gas puff from a nearby nozzle. The fibres are coupled to high sensitivity avalanche photo diodes and the signals are digitized at a rate of $2\times10^6$ frames per second. The in-focus spot size is $3.8\,\text{mm}$ for each of the individual fibre channels. The radial position of the last closed magnetic flux surface at the vertical centre of the image, $Z=-2.61\,\text{cm}$, is in the range from $89.4$ to $89.7\,\text{cm}$ for all the discharges presented here. The limiter radius mapped to this vertical position is at $R=91.0\,\text{cm}$. For each discharge the gas puff imaging diagnostic yields $0.25\,\text{s}$ usable data time series during the flat-top of the plasma current. By combining data from two discharges at the same $\nebar$ and two vertically adjacent views at the same radial position with identical statistical properties, time series of effectively one second duration are obtained. This allows calculation of statistical averages with high accuracy. Further information about the gas puff imaging diagnostic on Alcator C-Mod can be found in \Refs{cziegler,garcia-aps}.

\begin{figure}
\centering
\includegraphics[width=10cm]{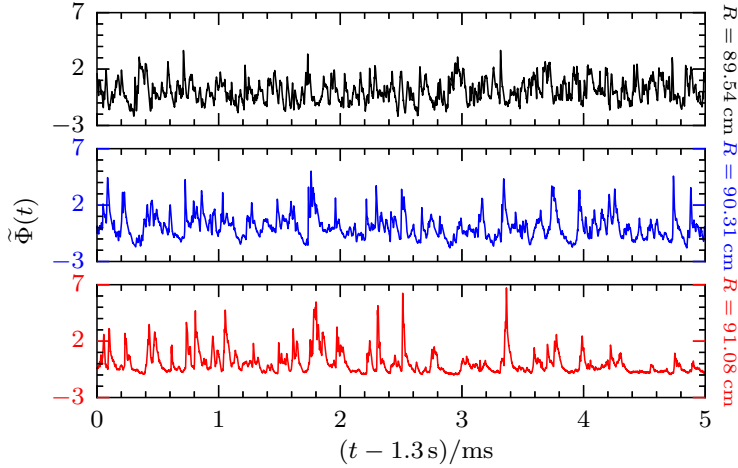}
\caption{Data time series recorded by the gas puff imaging diagnostic at the separatrix (top), in the main scrape-off layer (middle) and in the limiter shadow (bottom) for $\nebar/\nG=0.25$. All time series are rescaled such as to have vanishing mean and unit standard deviation.}
\label{fig:traw}
\end{figure}

Excitation of neutral He gas, and thus the intensity of the gas puff imaging signals, is determined by the neutral He particle density and a combination of the local electron density and temperature. This diagnostic does therefore not give information about the absolute value of any plasma parameters but is rather a proxy for the relative plasma fluctuation level. For this reason, the measurement data for each diode view position is here rescaled such as to have vanishing mean and unit standard deviation. Thus, a measured signal $\Phi(t)$ is normalized as $\wt{\Phi}=(\Phi-\overline{\Phi})/\overline{\Phi}_\text{rms}$, where $\overline{\Phi}$ and $\overline{\Phi}_\text{rms}$ are the moving average and standard deviation taken over a window of $8\,\text{ms}$ duration in order to remove low-frequency oscillations and trends in the signals. Previous analysis of Langmuir probe data have demonstrated that the statistical properties of the ion saturation current fluctuations in the far scrape-off layer are the same as that inferred from gas puff imaging data \cite{theodorsen-ppcf,garcia-aps,kube-ppcf,garcia-nme}. The major advantage of the gas puff imaging diagnostic for these studies is the access to long fluctuation data time series for several radial positions in the scrape-off layer \cite{cziegler,garcia-aps}.

As a representative example of the plasma fluctuations in the boundary region, gas puff imaging signals recorded at three different radial positions for $Z=-2.61\,\text{cm}$ and $\nebar/\nG=0.25$ is presented in \Figref{fig:traw}. The innermost diode view position shown here at $R=89.54\,\text{cm}$ is close to the separatrix and the outermost position at $R=91.08\,\text{cm}$ is in the limiter shadow where the magnetic field lines intersect limiter structures and have a connection length of the order of $1\,\text{m}$. It is clearly seen that the time series from the scrape-off layer are dominated by large-amplitude bursts, attributed to the radially outwards motion of blob-like plasma filaments. Further inward, pulses appear more frequently and at the separatrix the signal resembles random noise. However, the following analysis suggests that the underlying pulse structure and temporal scale are the same for all these time series, the only difference being the degree of pulse overlap.

Consider a stochastic process given by the super-position of a random sequence of $K$ pulses in a time interval of duration $T$ \cite{garcia-php,garcia-psd,theodorsen-ps,theodorsen-php},
\begin{equation}\label{shotnoise}
\Phi(t) = \sum_{k=1}^{K(T)} A_k\varphi( t-t_k ) ,
\end{equation}
where for each event $k$ the pulse amplitude is $A_k$ and the pulse arrival time is $t_k$, the latter assumed to be uniformly distributed on the interval $[0,T]$. The pulse shape is taken to be the same for all events and described by a two-sided exponential function \cite{garcia-psd,theodorsen-php}
\begin{equation}\label{pulse}
\varphi(t) = 
\begin{cases}
\exp{( t/\lambda\taud )}, & \quad t<0 ,
\\
\exp{( - t/(1-\lambda)\taud )}, & \quad t \geq 0 ,
\end{cases}
\end{equation}
where $\taud$ is the pulse duration and $\lambda$ is a pulse shape asymmetry parameter restricted to the range $0<\lambda<1$. For $\lambda<1/2$ the pulse function has a shorter rise than fall time.

In agreement with previous measurements on several tokamak experiments, it is in the following assumed that the pulse amplitudes are exponentially distributed with mean value $\Aave$ and the waiting times between the pulses are exponentially distributed with mean value $\tauw$ \cite{garcia-aps,theodorsen-ppcf,kube-ppcf,garcia-nme}. Introducing the scaled variable $\wt{\Phi}=(\Phi-\Phiave)/\Phirms$ with zero mean and unit standard deviation, similar to the experimental signals discussed above, the stationary probability density function for the random variable $\Phi$ is a Gamma distribution, giving $\wt{\Phi}$ the distribution \cite{garcia-php}
\begin{equation}\label{pgamma}
P_\wt{\Phi}(\wt{\Phi}) = \frac{\gamma^{1/2}}{\Gamma(\gamma)} \left( \gamma + \gamma^{1/2}\wt{\Phi} \right)^{\gamma-1} \exp\left( \gamma + \gamma^{1/2}\wt{\Phi} \right) ,
\end{equation}
where $\Gamma$ denotes the Gamma function. Here the shape parameter $\gamma=\taud/\tauw$ determines the degree of pulse overlap. It should be noted that since $\Phi>0$ it follows that $\wt{\Phi}>-\gamma^{1/2}$. The mean value of the signal given by \Eqref{shotnoise} is $\gamma\Aave$, the standard deviation is $\Phirms=\gamma^{1/2}\Aave$, and the skewness and flatness moments are $S_{{\Phi}}=2/\gamma^{1/2}$ and $F_{{\Phi}}=3+6/\gamma$, respectively. Accordingly, there is a parabolic relation between the skewness and flatness moments given by $F_{\Phi}=3+3S_{\Phi}^2/2$ \cite{garcia-php}. Like the relative fluctuation level given by $\Phirms^2/\Phiave^2=1/\gamma$, also the skewness and flatness moments increase with decreasing $\gamma$. The ratio $\gamma$ of pulse duration and waiting times is thus referred to as the \emph{intermittency parameter} of the model.  When $\gamma$ is small, pulses generally appear isolated in realizations of the process, resulting in a strongly intermittent signal. When $\gamma$ is large, there is significant overlap of pulses and realizations of the process resemble random noise and the probability density approaches as normal distribution. It should be noted that the pulse asymmetry parameter $\lambda$ does not enter the probability density function in \Eqref{pgamma} or any of its moments.

In the case of the two-sided exponential pulse shape given by \Eqref{pulse}, the frequency power spectral density can readily be calculated \cite{garcia-psd},
\begin{equation}\label{psd}
\Omega_{\wt{\Phi}}(\omega) = \frac{2\taud}{\left[ 1+(1-\lambda)^2(\taud\omega)^2 \right]\left[ 1+\lambda^2(\taud\omega)^2 \right]} ,
\end{equation}
where $\omega$ is the angular frequency. Thus, the spectrum is flat for low frequencies and has a steep power law shape for high frequencies. It should be noted that the intermittency parameter $\gamma$ does not influence the shape of the power spectral density, it is determined only by the pulse shape given in \Eqref{pulse} \cite{garcia-psd,garcia-php}.

The number of up-crossings $X_{\wt{\Phi}}$ of a given threshold level $\wt{\Phi}$ can be calculated in closed form for this stochastic process \cite{theodorsen-php,garcia-php}
\begin{equation}\label{xphi}
\frac{\taud}{T}\,X_{\wt{\Phi}}(\wt{\Phi}) = \frac{\lambda^{\gamma\lambda-1} \left( 1-\lambda \right)^{\gamma\left( 1-\lambda \right)-1}}{\gamma \Gamma\left( \gamma \lambda \right) \Gamma\left( \gamma \left( 1-\lambda \right) \right)} \left( \gamma + \gamma^{1/2}\wt{\Phi} \right)^\gamma \exp\left( - \gamma - \gamma^{1/2}\wt{\Phi} \right) .
\end{equation}
Here it should be noted that the pulse  asymmetry parameter $\lambda$ only appears in the pre-factor and therefore does not influence the functional form of the level crossing rate. The number of level crossings is evidently proportional to the duration $T$ of the process and is inversely proportional to the pulse duration time $\taud$. The rate of level crossings is highest for thresholds close to the mean value of the process. In the normal regime, $\gamma\gg1$, there are few crossings for threshold levels much smaller or much larger than the mean value due to the low probability of large-amplitude fluctuations. The rate of level crossings is therefore a narrow Gaussian function in this limit. In the strong intermittency regime, $\gamma\ll1$, the signal spends most of the time close to zero value, and virtually any pulse arrival will give rise to a level crossing for finite threshold values. The rate of level crossings therefore approaches a step function in this limit \cite{garcia-php,theodorsen-php}.

\begin{figure}
\centering
\includegraphics[width=10cm]{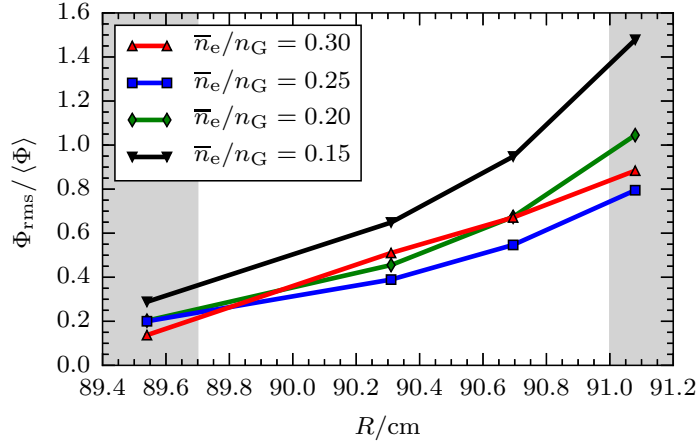}
\caption{Radial profile of the relative fluctuation level for four different line-averaged densities. The shaded region to the left indicates the position of the last closed magnetic flux surface, while the shaded region to the right indicates the limiter shadow.}
\label{fig:gamma}
\end{figure}

The radial variation of the relative fluctuation level, estimated by the ratio of the sample mean and standard deviation, for all four line-averaged densities is presented in \Figref{fig:gamma}. The relative fluctuation level is small close to the separatrix, indicating significant overlap of pulse structures as is clearly seen from the raw data time series in \Figref{fig:traw}. The relative fluctuation level increases radially outwards and is of order unity in the far scrape-off layer and the limiter shadow region. This implies a small value of the intermittency parameter, consistent with the distinct large-amplitude bursts observed in the time series in \Figref{fig:traw}.

\begin{figure}
\centering
\includegraphics[width=10cm]{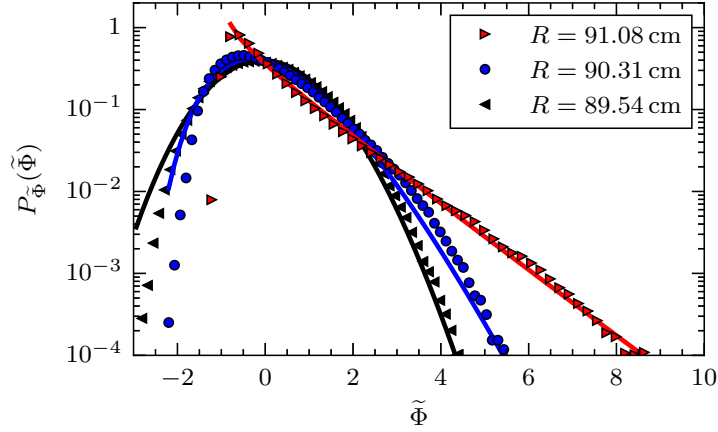}
\caption{Probability density functions of gas puff imaging data time series for $\nebar/\nG=0.25$ and various radial positions in the boundary region. The full lines are the best fits for predictions of the stochastic model for $\wt{\Phi}>2$.}
\label{fig:pdf}
\end{figure}

The probability density function of the gas puff imaging data for $\nebar/\nG=0.25$ is presented in \Figref{fig:pdf} for three different radial positions. The points show the estimated distribution from the detrended measurement data time series, while the solid lines show modified Gamma distributions given by \Eqref{pgamma}. The intermittency parameter is obtained from a least squares fit to the tail of the experimental distribution for amplitudes $\wt{\Phi}>2$. In the vicinity of the last closed magnetic flux surface, the fluctuations are close to normally distributed and the estimated intermittency parameter is close to 300. Radially outwards, the distribution becomes increasingly skewed and flattened, with $\gamma=11$ estimated at $R=90.31\,\text{cm}$. In the limiter shadow region, there is a clear exponential tail for large fluctuation amplitudes and the estimated intermittency parameter is slightly less than unity. The amplitude distribution for the gas puff imaging data are well described by predictions of the stochastic model, given by \Eqref{pgamma}, for all radial positions, except for small fluctuation amplitudes due to additional noise in the measurement data \cite{garcia-aps}. The addition of normal distributed noise to the variable $\Phi(t)$ does not significantly improve the model predictions for these data, thus only the tail behavior is considered for the amplitude distribution and level crossing rate in the following.

\begin{figure}
\includegraphics[width=10cm]{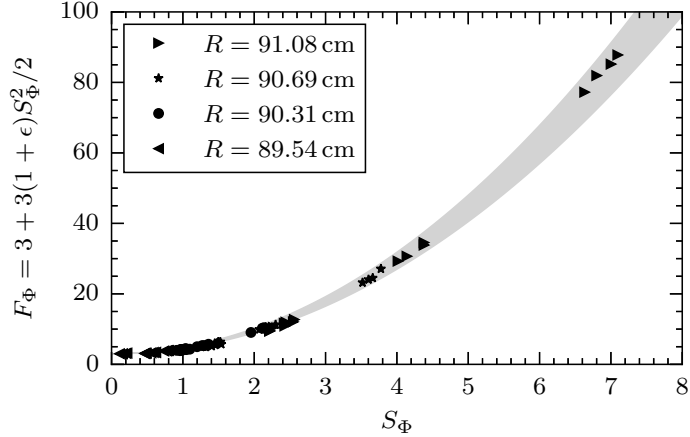}
\caption{Sample flatness versus skewness for all recorded gas puff imaging diode view positions in the scrape-off layer and all discharges in the density scan. The shaded area is the prediction of the stochastic model for a signal to noise ratio $1/\epsilon$ in the range from $0$ (lower boundary) to $5$ (upper boundary).}
\label{fig:fvss}
\end{figure}

In order to further demonstrate the intermittency of the fluctuations, the sample flatness moment is plotted against the sample skewness moment in \Figref{fig:fvss} for eight gas puff imaging diode view positions in the boundary region and all eight discharges in the density scan. The gray shaded region shows a parabolic relation between flatness and skewness with additive random noise included and the signal to noise ratio $1/\epsilon$ in the range from $0$ to $5$, which clearly suggests the significant noise levels in the far scrape-off layer and limiter shadow region \cite{garcia-aps,theodorsen-ps}. In agreement with the results in \Figsref{fig:gamma} and~\ref{fig:pdf}, the skewness and flatness moments both increase radially outwards in the scrape-off layer and their parabolic relation is in excellent agreement with predictions of the stochastic model.

\begin{figure}
\includegraphics[width=10cm]{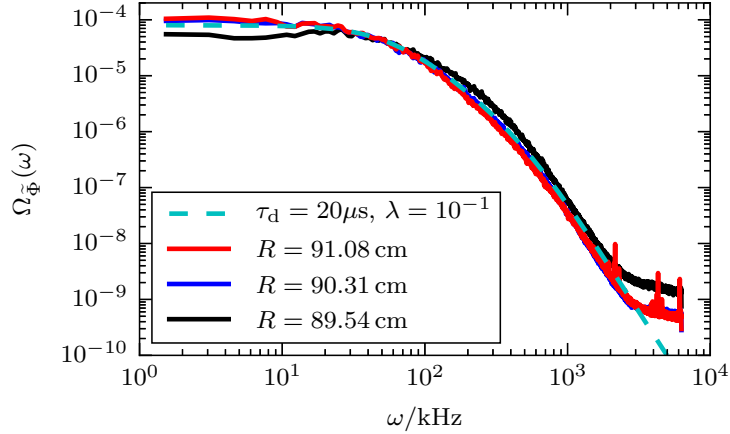}
\caption{Power spectral densities for the detrended gas puff imaging data time series for  $\nebar/\nG=0.25$ and various radial positions in the boundary region. The broken line is the power spectrum predicted by the stochastic model given by \Eqref{psd}.}
\label{fig:psd}
\end{figure}

\begin{figure}
\includegraphics[width=10cm]{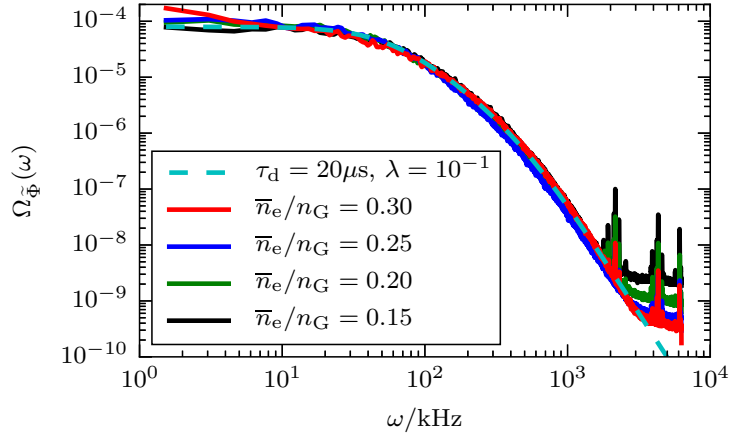}
\caption{Power spectral densities for the detrended gas puff imaging data time series for  $R=91.08\,\text{cm}$ and various line-averaged densities. The broken line is the power spectrum predicted by the stochastic model given by \Eqref{psd}.}
\label{fig:psdR}
\end{figure}

The power spectral densities for the gas puff imaging data at the last closed magnetic flux surface, the main scrape-off layer and the limiter shadow region are presented in \Figref{fig:psd} for $\nebar/\nG=0.25$. The remarkable similarity of these frequency spectra suggests that the fluctuations have the same temporal scale for all radial positions. The broken line in \Figref{fig:psd} is the prediction of the stochastic model given by \Eqref{psd} with $\taud=20\,\mu\text{s}$ and $\lambda=10^{-1}$. This is clearly an excellent description of the gas puff imaging data, except for the measurement noise that sets in at the highest frequencies. Similar results are found also for the other line-averaged densities, as is clearly seen in \Figref{fig:psdR} for $R=91.08\,\text{cm}$. The temporal scale is clearly comparable for all radial positions in the scrape-off layer and all line-averaged densities in these experiments. An analysis of the fluctuations measured radially inside the separatrix reveals significant variations of the auto-correlation function and power spectral density with the line-averaged density. This suggests that other processes than that described by the stochastic model given by \Eqref{shotnoise} may be present in the closed field line region.

\begin{figure}
\includegraphics[width=10cm]{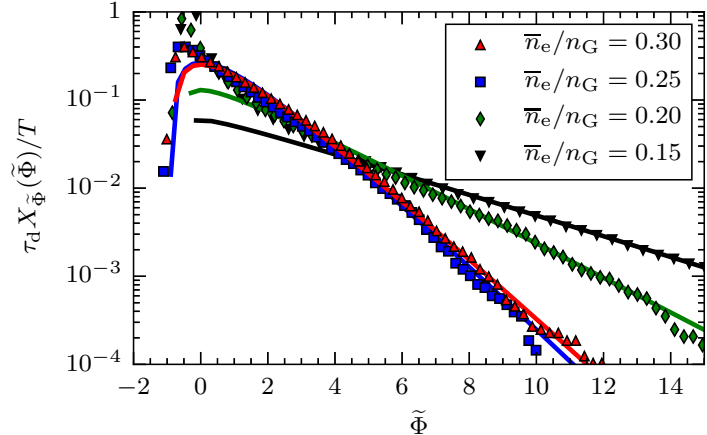}
\caption{Normalized rate of level crossings for the detrended gas puff imaging data time series at $R=91.08\,\text{cm}$ for various line-averaged densities. The full lines are the predictions of the stochastic model given by \Eqref{xphi}.}
\label{fig:xphi}
\end{figure}

The normalized rate of up-crossings of a given threshold level is presented in \Figref{fig:xphi} for the outermost diode view position and all four line-averaged densities. The full lines are the predictions of the stochastic model given by \Eqref{xphi}, where the intermittency parameter is taken from the fit of a Gamma distribution to the tail of the probability density function, as described in connection to \Figref{fig:pdf}. In all cases, the pulse duration is set to $\taud=20\,\mu\text{s}$ and the pulse asymmetry parameter is set to $10^{-1}$, consistent with the power spectral densities presented in \Figref{fig:psdR}. This compares favorably with the tail behavior for the experimental measurements for all line-averaged densities. As expected, the model fails to describe the high rate of level crossings for small fluctuation amplitudes due to the significant measurement noise at this radial position.

The predictions of a stochastic model are here shown to reproduce many of the statistical properties of the gas puff imaging measurements on Alcator C-Mod. This includes the probability density function, the power spectral density and the rate of threshold crossings. Previous investigations have demonstrated that the assumptions underlying the model are supported by the measurement data, namely a super-position of uncorrelated exponential pulses with an exponential distribution of pulse amplitudes \cite{garcia-aps,theodorsen-ppcf,kube-ppcf,garcia-nme}.

\begin{figure}
\includegraphics[width=10cm]{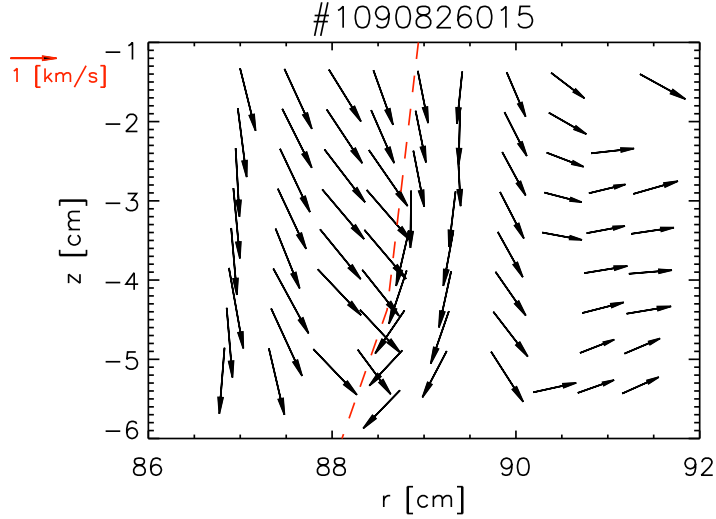}
\caption{Two-dimensional velocity field of the edge fluctuations based on a cross-correlation analysis. The vertical dashed line is the position of the last closed magnetic flux surface. The reference arrow on the top left of the figure corresponds to a magnitude of $1\,\text{km/s}$. Reproduced with permission from \Ref{agostini}.}
\label{fig:agostini}
\end{figure}

The analysis presented here clearly demonstrate that the plasma fluctuations in the boundary region of Alcator C-Mod become increasingly intermittent with radial distance into the scrape-off layer. Despite this, there is a remarkable similarity of the frequency power spectra for all radial positions and line-averaged densities investigated. Based on the predictions of the stochastic model, this implies that the temporal scale for the fluctuations is the same in all cases. Near the last closed flux surface, there is significant overlap of pulse structures leading to normally distributed fluctuation amplitudes and low intermittency. In the far scarpe-off layer, the pulses appear separated and intermittency is pronounced. The frequency spectra are entirely determined by the shape of the underlying pulses \cite{garcia-psd}.

The much stronger degree of pulse overlap in the vicinity of the last closed magnetic flux surface is likely due to the dominant poloidal flow in the edge and near scrape-off layer regions. An example of this is presented in \Figref{fig:agostini}, which is based on gas puff imaging by a $64\times64$ pixel Phantom fast-framing camera in a plasma with similar conditions as those investigated above \cite{agostini}. The camera field of view is divided into smaller square regions of $10\times10$ pixels, and the cross-correlation between the pixels in each square is evaluated. From the time delay of the maximum of the correlation it is possible to compute the two-dimensional velocity field. This clearly shows a dominantly poloidal velocity in the vicinity of the last closed magnetic flux surface and a dominantly radial velocity in the far scarpe-off layer \cite{agostini,terry}.

Moreover, some of the blob-like structures that propagate through the scrape-off layer will likely be subject to non-linear dispersion and break up, resulting in small-amplitude noise which enhances the rate of level crossings for small threshold values \cite{garcia-blob,kube-blob}. However, most large-amplitude structures are expected to propagate through the scrape-off layer and into the limiter shadow, resulting in an exponential tail in the probability density function and frequent crossings of large threshold levels.

Based on the rate of level crossings, the stochastic model promises to be a valuable tool for predicting threshold phenomena like plasma--wall interactions due to transient transport events \cite{garcia-php,theodorsen-php}. The absolute rate of level crossings depends on the intermittency parameter $\gamma$, the pulse duration time $\taud$ and the mean value $\Phiave$ of the plasma parameter under consideration. The latter cannot be inferred from gas puff imaging data, and thus motivates a systematic investigation of the fluctuation statistics using the novel mirror Langmuir probe system implemented on Alcator C-Mod \cite{labombard-mlp}. Moreover, the analysis presented here will be extended to include measurement data from high performance plasmas and improved confinement modes. Finally, a systematic multi-machine comparison should be made in order to confirm the universality of the fluctuations and reveal any dependence of the model parameters on for example machine size.

\section{Acknowledgements}

This work was supported with financial subvention from the Research Council of Norway under grant 240510/F20 and the U.S. Department of Energy, Office of Science, Office of Fusion Energy Sciences, using User Facility Alcator C-Mod, under Award Number DE-FC02-99ER54512-CMOD. A.~T.\ and O.~E.~G.\ acknowledge the generous hospitality of the MIT Plasma Science and Fusion Center where this work was conducted.

\end{document}